\documentstyle[12pt]{article}
\newcommand{\bea}{\begin{eqnarray}}
\newcommand{\eea}{\end{eqnarray}}
\newcommand{\be}{\begin{equation}}
\newcommand{\ee}{\end{equation}}
\newcommand{\vs}[1]{\vspace{#1 mm}}

\renewcommand{\a}{\alpha}
\renewcommand{\b}{\beta}

\renewcommand{\d}{\delta}

\newcommand{\dsl}{\pa \kern-0.5em /}

\newcommand{\pa}{\partial}

\newcommand{\nn}{\nonumber\\}

\begin{document}
\topmargin 0pt
\oddsidemargin 0mm

\begin{flushright}
USTC-ICTS-05-6\\
hep-th/0506115\\
\end{flushright}

\vs{2}
\begin{center}
{\Large \bf Tachyon condensation on the intersecting
brane-antibrane system} \vs{10}

{\large Hua Bai$^a$\footnote{E-mail: huabai@mail.ustc.edu.cn}, J.
X. Lu$^{a,b,c}$\footnote{E-mail: jxlu@ustc.edu.cn}
 and Shibaji Roy$^d$\footnote{E-mail: roy@theory.saha.ernet.in}}

 \vspace{5mm}

{\em
 $^a$ Interdisciplinary Center for Theoretical Study\\
 University of Science and Technology of China, Hefei, Anhui 230026,
P. R. China\\

\vs{4}

$^b$ Center for Mathematics and Theoretical Physics\\
Shanghai Institute for Advanced Study\\
University of Science and Technology of China, Shanghai 201315, P.
R.  China\\

\vs{4}

$^c$ Interdisciplinary Center of Theoretical Studies\\
Chinese Academy of Sciences, Beijing 100080, P. R. China\\

\vs{4}

 $^d$ Saha Institute of Nuclear Physics,
 1/AF Bidhannagar, Calcutta-700 064, India}
\end{center}

\vs{5}
\centerline{{\bf{Abstract}}}
\vs{5}
\begin{small}
We generalize our study of the tachyon condensation on the
brane-antibrane system [hep-th/0403147] to the intersecting
brane-antibrane system. The supergravity solutions of the
intersecting brane-antibrane system are characterized by five
parameters. We relate these parameters to the microscopic physical
parameters, namely, the number of D$p$-branes ($N_1$), the number
of ${\bar {\rm D}}p$-branes (${\bar N}_1$), the number of
D$(p-4)$-branes ($N_2)$, the number of ${\bar{\rm D}}(p-4)$-branes
(${\bar N}_2$) and the tachyon vev $T$. We show that the solution
and the ADM mass capture all the required properties and give a
correct description of the tachyon condensation for the
intersecting brane-antibrane system.
\end{small}
\newpage

A coincident D-brane--antiD-brane pair (or a non-BPS D-brane) in
Type II string theories is known to be unstable which is
characterized, for example, by the presence of tachyonic mode on
the D-brane world-volume \cite{asone}. As a result, these systems
decay and the decay occurs by a process known as tachyon
condensation \cite{astwo}. Tachyon condensation is well understood
in the open string description using either the string field
theory approach \cite{sz,gs} or the tachyon effective action
approach \cite{kl} on the brane. In \cite{luroyone} we have
obtained a closed string (or supergravity) understanding of this
process. We have interpreted \cite{bmo} the previously known
\cite{zz,im} non-supersymmetric, three parameter supergravity
solutions with a symmetry ISO($p,1$) $\times$ SO($9-p$) in ten
space-time dimensions as the coincident D$p$--${\bar {\rm D}}p$
system. Then the three parameters in this solution were related to
the microscopic physical parameters, namely, the number of
D$p$-branes ($N$), the number of ${\bar {\rm D}}p$-branes (${\bar
N}$) and the tachyon vev of the D$p$--${\bar {\rm D}}p$ system.
Using these relations we have calculated the ADM mass and have
shown that the solution and the ADM mass capture all the required
properties and give a correct description of the tachyon
condensation advocated by Sen \cite{astwo} on the D-${\bar{\rm
D}}$ system.

In this paper we generalize our previous study \cite{luroyone} of
the tachyon condensation on the brane-antibrane system to the
intersecting brane-antibrane system. The supergravity solution in
this case is again non-supersymmetric and has the isometry
ISO($p-4,1$) $\times$ SO(4) $\times$ SO($9-p$) which can obviously
be identified as the intersecting brane-antibrane system or more
precisely the intersecting D$p$-${\bar{\rm D}}p$ and
D$(p-4)$-${\bar {\rm D}}(p-4)$ system for $3\leq p \leq 6$. We
find that the solution is characterized by five parameters and
once the solution is realized to represent intersecting
D$p$-${\bar{\rm D}}p$ and D$(p-4)$-${\bar {\rm D}}(p-4)$ system,
we can immediately infer that the parameters must be related to
the microscopic physical parameters $N_1$, the number of
D$p$-branes, ${\bar N}_1$, the number of $\bar{\rm D}p$-branes,
$N_2$, the number of D$(p-4)$-branes, ${\bar N}_2$, the number of
$\bar{\rm D}(p-4)$-branes and the tachyon vev $T$\footnote{So long
as the bulk configuration is concerned, the worldvolume fields (in
particular the tachyon) don't need to satisfy their respective
worldvolume equations of motion (for example, we can put
worldvolume scalars and tachyon to constants and other fields to
zero) in the spirit, for example, of \cite{bmo}. In other words,
they can be off-shell. In this way, the tachyon vev will appear as
a parameter labelling the solution.}. The more exact relationships
between the supergravity parameters and the microscopic physical
parameters can be obtained by examining how the solution reduces
to the supersymmetric configuration which corresponds to the four
cases as follows. (a) The intersecting $N_1$ D$p$-branes and $N_2$
D$(p-4)$-branes (when both ${\bar N}_1$ and ${\bar N}_2$ are
zero). If in addition to ${\bar N}_1 = {\bar N}_2 = 0$ we also
have $N_2=0$ (or $N_1=0$), we still get a BPS configuration of
half susy instead of quarter susy (as is the case for the
intersecting D$p$/D$(p-4)$ brane configuration) which is $N_1$
D$p$-branes (or delocalized $N_2$ D$(p-4)$-branes).  When all the
$N$'s are zero, which is a trivial case, we get the maximally
supersymmetric flat space-time. (b) The intersecting ${\bar N}_1$
${\bar{\rm D}}p$-branes and ${\bar N}_2$ ${\bar {\rm
D}}(p-4)$-branes (when both $N_1$ and $N_2$ are zero). Similar to
the previous case here also if in addition to $N_1=N_2=0$, we have
${\bar N}_2=0$ (or ${\bar N}_1=0$) we get half susy configuration
which is  ${\bar N}_1$ ${\bar {\rm D}}p$-branes (or delocalized
${\bar N}_2$ ${\bar {\rm D}}(p-4)$-branes). (c) The intersecting
$N_1$ D$p$-branes and ${\bar N}_2$ ${\bar {\rm D}}(p-4)$-branes
(when both ${\bar N}_1$ and $N_2$ are zero). This case actually
preserves also one quarter of spacetime supersymmetry which seems
impossible from a first look\footnote{The ($p, p'$) system of
D-branes with $p - p' = 4$ is very special and when one calculates
the one-loop interaction amplitude between them (for example, see
\cite{dvl}), the R-R amplitude has zero contribution while the
NS-NS amplitude has two terms cancelled for this case. Since the
NS-NS amplitude is insensitive to the sign of the charges carried
by the individual D-brane,  therefore the force acting between the
two branes vanishes, independent of the sign of their charges.
Such a no-force condition indicates that there is a certain
fraction of spacetime supersymmetry preserved. Further when two
kinds of D-branes with different dimensionality intersect, the
spacetime supersymmetry preserved is determined by the following
two equations: $\epsilon_1 = \eta \Gamma^0 \Gamma^1 \cdots
\Gamma^p \,\epsilon_2$ and $\epsilon_1 = \eta' \Gamma^0 \Gamma^1
\cdots \Gamma^{p'}\, \epsilon_2$ with $\Gamma^\mu$ the ten
dimensional $\gamma$-matrix, and $\epsilon_1$ and $\epsilon_2$ the
two Majorana-Weyl supersymmetry parameters in IIA/IIB string
theory (In IIA, $p$ and $p'$ are both even while in IIB they are
odd). Here $\eta = \pm$ and $\eta' = \pm $ label the sign of the
charge carried by the corresponding branes, respectively. Assuming
$p > p'$ and defining $\Gamma = \Gamma^{p' + 1} \cdots  \Gamma^p$,
one can show that only for $\Gamma^2 = I$ with $I$ the unit
matrix, i.e., for, $(- 1)^{(p - p')(p - p' + 1)/2} = 1$, a quarter
of susy is preserved, independent of the values of $\eta$ and
$\eta'$. In a given theory, $p - p'$ is even and the above
condition implies $p - p' = 4, 8$. As $p = p'$, $\Gamma$ is a unit
matrix, then only for $\eta = \eta'$, one half of susy can be
preserved. Here we discuss the so-called threshold bound states.
For nonthreshold bound state, one half of susy can be preserved
for $p - p' = 2$ based on U-duality.}. Again similar to the
previous case here also if in addition to ${\bar N}_1= N_2=0$, we
have ${\bar N}_2=0$ (or $ N_1=0$) we get half susy configuration
which is $N_1$ D$p$-branes (or delocalized ${\bar N}_2$ ${\bar
{\rm D}}(p-4)$-branes). (d) This is similar in sprit to the case
(c) but now with ${\bar N}_1$ and $N_2$ non-zero. Other special
cases can be discussed accordingly.

 When none
of the $N$'s are zero, in general, the solution is not
supersymmetric and there is a tachyon on the world-volume of the
intersecting branes\footnote{The tachyon can be expressed
as
$T = \left(\begin{array}{cc}T_1& T_2\\
                              T_3 & T_4\end{array}\right)$.
 Here
$T_1$ is in the bi-fundamental $(N_1, {\bar N}_1)$ of gauge group
$U(N_1)\times U({\bar N}_1)$ , $T_2$ in $(N_1, {\bar N}_2)$ of
gauge group $U(N_1)\times U({\bar N}_2)$, $T_3$ in  $(N_2, {\bar
N}_1)$ of gauge group $U(N_2)\times U({\bar N}_1)$ and $T_4$ in
$(N_2, {\bar N}_2)$ of gauge group $U(N_2)\times U({\bar N}_2)$.
The tachyon used in the discussion is actually $\left(Tr\,(T{\bar
T})\right)^{1/2}$ with a proper normalization.}.  However we can
always get BPS configuration at the end of tachyon condensation
given the above discussion and the one in footnote 5. When $N_1 =
{\bar N}_1$ or $N_2 = {\bar N}_2$, we will restore one half of
susy  at the end of tachyon condensation. When the two hold
simultaneously, we end up with a flat spacetime at the end of
tachyon condensation with maximal supersymmetry. For all the other
cases, one quarter of susy is preserved at the end of tachyon
condensation. For example, when $N_1 > {\bar N}_1$ and $N_2>{\bar
N}_2$ we expect to have intersecting $(N_1-{\bar N}_1)$
D$p$-branes with $(N_2-{\bar N}_2)$ D$(p-4)$-branes at the end of
tachyon condensation which is supersymmetric (quarter BPS), on the
other hand, when $N_1={\bar N}_1$ we expect to have delocalized
$(N_2-{\bar N}_2)$ D$(p-4)$-branes at the end of tachyon
condensation which is half BPS and so on. The recognition for
having a supersymmetric background at the end of tachyon
condensation is crucial and we use all these information to relate
the supergravity parameters to the microscopic physical
parameters.

Now in order to understand the tachyon condensation, we look at the
expression of the total ADM mass of the intersecting brane-antibrane
supergravity solution representing the
total energy of the system. We then express this total energy in
terms of the five microscopic physical parameters namely, $N_1$, ${\bar N}_1$,
$N_2$, ${\bar N}_2$ and
$T$ of the intersecting D$p$-${\bar{\rm D}}p$ and D$(p-4)$-${\bar{\rm D}}(p-4)$
system using the aforementioned
relations. The total energy can be seen to be equal to or less than
the sum of the masses of $N_1$ D$p$-branes, ${\bar N}_1$ ${\bar
{\rm D}}p$-branes, $N_2$ D$(p-4)$-branes and ${\bar N}_2$ ${\bar{\rm D}}(p-4)$-branes
indicating the presence of tachyon contributing
the negative potential energy to the system. We will see that the
energy expression gives the right picture of tachyon condensation
as is expected of an intersecting brane-antibrane system. We will
reproduce all the
expected results from this general mass formula under various
special limits at
the top and at the bottom of the tachyon potential. We will also
show how the various known BPS supergravity configurations can be
reproduced in these special limits.

The non-supersymmetric intersecting D$p$/D$(p-4)$ supergravity
solution having isometry ISO($p-4,1$) $\times$ SO(4) $\times$
SO($9-p$) which can also be identified as the intersecting
D$p$-${\bar{\rm D}}p$ and D$(p-4)$-${\bar {\rm D}}(p-4)$ has the
form in space-time dimension $d=10$ as\footnote{This configuration
has also been considered previously in \cite{mo} but in a rather
different notations.}, \bea ds^2 &=& F_1^{-\frac{7-p}{8}}
F_2^{-\frac{11-p}{8}} \left(-dt^2 + \sum_{i=1}^{p-4}dx_i^2\right)
+ F_1^{-\frac{7-p}{8}} F_2^{\frac{p-3}{8}}
\sum_{j=p-3}^{p}dx_j^2\nn & & \qquad\qquad\qquad\qquad\qquad\qquad
+ \left(H\tilde{H}\right)^{\frac{2}{7-p}}F_1^{\frac{p+1}{8}}
F_2^{\frac{p-3}{8}} \left(dr^2 + r^2 d\Omega_{8-p}^2\right)\nn
e^{2\phi} &=& F_1^{\frac{3-p}{2}}
F_2^{\frac{7-p}{2}}\left(\frac{H}{\tilde{H}} \right)^{2\delta}\nn
F_{[8-p]} &=&
 b \,{\rm Vol}(\Omega_{8-p})\nn
F_{[12-p]} &=& c \,{\rm Vol}(\Omega_{8-p}) \wedge dx_{p-3} \wedge
\ldots \wedge dx_p \eea Note that in the above we have written the
metric in the Einstein frame. The Vol($\Omega_{8-p}$) represents
the volume form of the unit $(8-p)$-dimensional sphere. Also the
various functions appeared in the solution are defined below, \bea
F_1 &=& \cosh^2\theta_1
\left(\frac{H}{\tilde{H}}\right)^{\alpha_1} - \sinh^2\theta_1
\left(\frac{\tilde{H}}{H}\right)^{\beta_1}\nn F_2 &=&
\cosh^2\theta_2 \left(\frac{H}{\tilde{H}}\right)^{\alpha_2} -
\sinh^2\theta_2 \left(\frac{\tilde{H}}{H}\right)^{\beta_2}\nn H
&=& 1 + \frac{\omega^{7-p}}{r^{7-p}}, \qquad
\tilde{H}\,\,\,=\,\,\, 1 - \frac{\omega^{7-p}}{r^{7-p}} \eea Here
$\a_1$, $\b_1$, $\a_2$, $\b_2$, $\d$ and $\omega$ are integration
constants. $b$ and $c$ are the charge parameters related to the
net charges of the D$p$-${\bar{\rm D}}p$ and D$(p-4)$-${\bar {\rm
D}}(p-4)$-brane systems respectively. However, not all the
parameters are independent and they are related as, \bea \a_1 -
\b_1 &=& \frac{p-3}{2}\delta, \qquad b \,\,\,=\,\,\, (7-p)
(\a_1+\b_1) \omega^{7-p} \sinh2\theta_1\nn \a_2 - \b_2 &=&
\frac{p-7}{2}\delta, \qquad c \,\,\,=\,\,\, (7-p) (\a_2+\b_2)
\omega^{7-p} \sinh2\theta_2 \eea and from the consistency of
equations of motion we also get \be (\a_1+\b_1)^2+ (\a_2+\b_2)^2 +
\left(4-\frac{(p-3)^2}{4} -\frac{(p-7)^2}{4}\right) \d^2 = 8
\frac{8-p}{7-p} \ee So, treating $(\a_1+\b_1) \equiv \Delta_1$ and
$(\a_2+\b_2) \equiv \Delta_2$ as independent, there are actually
five independent parameters in the solution, namely, $\theta_1$,
$\theta_2$, $\omega$, $\Delta_1$ and $\Delta_2$. We will relate
each of these parameters to the physical microscopic parameters of
the intersecting D$p$-${\bar{\rm D}}p$ and D$(p-4)$-${\bar{\rm
D}}(p-4)$ system. But before we do that we would like to make some
comments. First of all, we mention that as $r \to \infty$, both
$H$ and ${\tilde H}$ $\to$ 1, and so, $F_1,\, F_2\, \to 1$. The
solution is therefore asymptotically flat. Also, note that the
solution (1) represents magnetically charged intersecting
brane-antibrane system. To obtain the electrically charged ones we
just make a transformation $F_{[8-p]} \to e^{\frac{3-p}{2}\phi}
\ast F_{[8-p]}$ and $F_{[12-p]} \to e^{\frac{7-p}{2}\phi}\ast
F_{[12-p]}$, where $\ast$ represents the Hodge dual. Using these
the gauge fields for the electrically charged solution can be
written as, \bea A_{[p+1]} &=& \sinh\theta_1 \cosh\theta_1
\left(\frac{C_1}{F_1}\right) dt\wedge dx_1 \wedge\ldots\wedge
dx_{p}\nn A_{[p-3]} &=& \sinh\theta_2 \cosh\theta_2
\left(\frac{C_2}{F_2}\right) dt\wedge dx_1 \wedge\ldots\wedge
dx_{p-4} \eea where $C_1$ and $C_2$ are defined as, \bea C_1 &=&
\left(\frac{H}{\tilde H}\right)^{\a_1} - \left(\frac{\tilde
H}{H}\right)^{\b_1}\nn C_2 &=& \left(\frac{H}{\tilde
H}\right)^{\a_2} - \left(\frac{\tilde H}{H}\right)^{\b_2} \eea We
note from (2) that the solution has a potential singularity at
$r=\omega$ and we will work only in the physically relevant region
$r>\omega$. Also, without any loss of generality we will choose
all of the parameters $\Delta_1$, $\Delta_2$ to be $\geq 0$. We
remark that the solution is non-supersymmetric can be seen from
the $(H\tilde H)^{2/(7-p)}$ factor in the last term of the metric
in eq.(1). This is consistent with our interpretation of the
solution to be intersecting brane-antibrane system. We can now
express the parameter $b$ and $c$ in terms of the number of
branes-antibranes using eq.(1) as follows, \bea Q_0^p\, (N_1 -
{\bar N}_1) &=& \frac{b \Omega_{8-p}}{\sqrt{2} \kappa_0} \quad
\Rightarrow \quad b\,\,\,=\,\,\, \frac{\sqrt{2}\kappa_0 Q_0^p
(N_1-{\bar N}_1)}{\Omega_{8-p}}\nn Q_0^{p-4}\, (N_2 - {\bar N}_2)
&=& \frac{c \Omega_{8-p} V_4}{\sqrt{2} \kappa_0} \quad \Rightarrow
\quad c\,\,\,=\,\,\, \frac{\sqrt{2}\kappa_0 Q_0^{p-4} (N_2-{\bar
N}_2)}{\Omega_{8-p} V_4} \eea where $Q_0^p =
(2\pi)^{(7-2p)/2}\alpha'^{(3-p)/2}$ is the unit charge on the
D$p$-brane and similarly $Q_0^{p-4}$ is the unit charge on the
D$(p-4)$-brane. $\sqrt{2}\kappa_0 = (2\pi)^{7/2} \alpha'^2$ is
related to 10 dimensional Newton's constant. $V_4$ is the compact
volume of the four directions $x_{p-3}$ to $x_p$. $\Omega_n = 2
\pi^{(n+1)/2}/ \Gamma((n+1)/2)$. Note that $b \to 0$ as $N_1 \to
{\bar N}_1$ and $c \to 0$ as $N_2 \to {\bar N}_2$ as expected.

For the solution (1), the supersymmetry will be restored if and
only if $H{\tilde H} \to 1$ which always requires $\omega^{7 -p}
\to 0$. As we have already stated in the beginning we have the
following cases for which supersymmetry will be restored: (1) $N_1
= N_2 = 0$, or, ${\bar N}_1 = {\bar N}_2 = 0$, or, $N_1 = {\bar
N}_2 = 0$, or, ${\bar N}_1 = N_2 = 0$  (or, $N_1 = N_2 = {\bar
N}_1 = {\bar N}_2 = 0$ which is the trivial case). (2) When none
of the groups are zero, we can still have supersymmetric
configuration at the end of tachyon condensation for which the
number of susy restored depends on the initial values of $N_1,
N_2, {\bar N}_1$ and ${\bar N}_2$ as discussed earlier. For
example, when $N_1 = {\bar N}_1$ and $N_2 = {\bar N}_2$, we expect
to get an empty space-time at the end of tachyon condensation with
maximal supersymmetry. What is important is the recognition that
for all these cases, $\omega^{7-p} \to 0$. This observation will
be crucial to relate $\omega$ as well as other parameters in terms
of the physical microscopic parameters of the brane-antibrane
system. We also point out that at the top of the tachyon potential
when $\omega^{7-p}$ does not go to zero, it must give the correct
number of branes so that the mass formula can be correctly
reproduced. All these information can be captured in a single
formula for $\omega^{7-p}$ as follows, \be (7-p) \omega^{7-p}\,\,
=\,\, \sqrt{\frac{7-p}{2(8-p)}}
\frac{2\kappa_0^2}{\Omega_{8-p}}T_p \left[\sqrt{N_1 {\bar N}_1} +
a \sqrt{N_2{\bar N}_2}\right]
 \cos T
\ee
 In the above $T_p =
(2\pi)^{-p}\alpha'^{-(p+1)/2}$ is the tension of a D$p$-brane. The
factor $2\kappa_0^2 T_p/\Omega_{8-p}$ in front is kept so that it
will give the correct mass of the system. Also we have defined
$a=T_{p-4}/(V_4 T_p) = (2\pi \sqrt{\alpha'})^4/V_4$ relating the
tensions of a D$p$-brane and a D$(p-4)$-brane. We also point out
that $T$ in the above represents the tachyon vev and we have taken
$T=0$ as the top of the tachyon potential and $T=\pi/2$ as the
bottom of the potential. Thus all the cases we have discussed
above is contained in $\omega^{7-p}$ in eq.(8).

Now having obtained the form of $\omega^{7-p}$ in terms of the
microscopic physical parameters, we would like to obtain similar
relations for the other parameters as well. But first we will try
to relate $\delta$, $\Delta_1$ and $\Delta_2$ in terms of the
microscopic parameters. But before that we observe that the total
ADM mass of the intersecting D$p$-${\bar {\rm D}}p$ and
D$(p-4)$-${\bar {\rm D}}(p-4)$ can be obtained from the
supergravity configuration (1) using the formula given in
\cite{jxl} as, \be M = \frac{\Omega_{8 - p}}{2 \kappa^2_0} (7 - p)
\omega^{7 - p} \left[\left(\Delta_1 \cosh2\theta_1 + \frac{p-3}{2}
\delta\right) + \left(\Delta_2 \cosh2\theta_2 + \frac{p-7}{2}
\delta\right)\right] \ee This clearly shows that the ADM mass or
the total energy of the system breaks up into two parts
corresponding to the D$p$-${\bar {\rm D}}p$ and the
D$(p-4)$-${\bar {\rm D}}(p-4)$-brane systems. From our earlier
experience on the tachyon condensation on the brane-antibrane
system we find that in this case the parameters $\Delta_1$ and
$\Delta_2$ satisfy the following relations in terms of $\delta$
\bea \Delta_1^2 &=& \frac{(p-3)^2}{4} \delta^2 -
\frac{(p-3)\d}{\gamma A} \sqrt{\frac{(N_1 - {\bar N}_1)^2}{\cos^2
T} + 4 N_1 {\bar N}_1 \cos^2T} + \frac{4N_1 {\bar N}_1\cos^2
T}{\gamma^2 A^2}\nn \Delta_2^2 &=& \frac{(p-7)^2}{4} \delta^2 -
\frac{(p-7)a\d}{\gamma A} \sqrt{\frac{(N_2 - {\bar N}_2)^2}{\cos^2
T} + 4 N_2 {\bar N}_2 \cos^2T} + \frac{4 a^2 N_2 {\bar N}_2\cos^2
T}{\gamma^2 A^2} \eea where $\gamma=\sqrt{(7-p)/2(8-p)}$,
$\Delta_1 = \a_1+ \b_1$, $\Delta_2 = \a_2+\b_2$ and $A =
\sqrt{N_1{\bar N}_1} + a \sqrt{N_2{\bar N}_2}$. Then using the
parameter relation (4) we find that the parameter $\d$ satisfy the
following quadratic relation where only the $\d < 0$ root will be
relevant for our discussion \bea & &\d^2 - \frac{\d}{4\gamma A}
\left[(p-3) \sqrt{\frac{(N_1-{\bar N}_1)^2}{\cos^2 T} + 4 N_1
{\bar N}_1 \cos^2 T} + a (p-7) \sqrt{\frac{(N_2-{\bar
N}_2)^2}{\cos^2 T} + 4 N_2 {\bar N}_2 \cos^2 T}\right]\nn & & -
\frac{(N_1{\bar N}_1  + a^2 N_2 {\bar N}_2) \sin^2 T + 2 a
\sqrt{N_1 N_2 {\bar N}_1 {\bar N}_2} }{\gamma^2 A^2}\,\,\,=\,\,\,0
\eea Thus $\d$ is given entirely in terms of the microscopic
physical parameters. As $T\to 0$, the $\d$ approaches a finite
negative value while as $T\to \pi/2$, $\d = - 4 (A/\gamma) \cos
T/[(p -3) |N_1 - {\bar N}_1| + a (p - 7) |N_2 - {\bar N}_2|]$ if
not all the four $N_1, N_2, {\bar N}_1, {\bar N}_2$ are equal and
if so, then $\d = - 1/\gamma$. Once $\d$ is known we can determine
$\Delta_1$ and $\Delta_2$ in terms of the microscopic parameters
as well. After we determine the forms of $\Delta_1$ and $\Delta_2$
then using the form of $\omega^{7-p}$ given in (8) we can easily
relate the parameters $\theta_1$ and $\theta_2$ to the microscopic
physical parameters using (3) and (7) as, \bea \sinh2\theta_1 &=&
\frac{|N_1-{\bar N}_1|}{\gamma\Delta_1 A \cos T },\nn
\sinh2\theta_2 &=& \frac{a|N_2-{\bar N}_2|}{\gamma\Delta_2 A \cos
T }, \eea where we have assumed both $\theta_1$ and $\theta_2$ to
be $\geq 0$.
 We note from (12) that as $N_1 \to {\bar N}_1$,
$\theta_1 \to 0$. We have also seen it before from eq.(7) that
$N_1 \to {\bar N}_1$ implies $b \to 0$ and so, $\theta_1 \to 0$
implies the charge $b \to 0$. Similarly, $\theta_2 \to 0$ implies
the charge $c \to 0$. Note from eq.(4) that the values of
$|\Delta_1|, |\Delta_2|$ and $|\d|$ are all bounded from the
above, therefore from (12) we see that both $\theta_1$ and
$\theta_2$ blow up at the end of the tachyon condensation.

Now all the quantities in the ADM mass expression given in eq.(9)
are known and so, substituting $\omega^{7-p}$, $\Delta_1$,
$\Delta_2$ and $\d$ we find that the mass expression has a very
simplified form given by \bea M &=& T_p \left[\sqrt{(N_1-{\bar
N}_1)^2 + 4 N_1 {\bar N}_1 \cos^4 T} + a \sqrt{(N_2-{\bar N}_2)^2
+ 4 N_2 {\bar N}_2 \cos^4 T}\right]\nn & \leq & T_p
\left[(N_1+{\bar N}_1) + a (N_2 + {\bar N}_2)\right] \eea
 We thus
note that the total mass per unit $p$-brane volume of the system
is less or equal to the sum of the those of $N_1$ D$p$-branes,
${\bar N}_1$ ${\bar {\rm D}}p$-branes, $N_2$ D$(p-4)$-branes,
${\bar N}_2$ ${\bar {\rm D}}(p-4)$-branes and the difference is
the tachyon potential energy per unit $p$-brane volume $V(T)$
which is negative. We can easily check that $T=0$ gives the
maximum of the energy, therefore the maximum of the tachyon
potential (which is zero) while $T=\pi/2$ gives the corresponding
minima. We want to point out that eq.(13) is consistent with our
previous experience for a simple brane-antibrane system discussed
in \cite{luroyone} and the special feature for the present system
under consideration is that no interaction exists between two
D-branes with their dimensionality differing by four as discussed
in footnote 5.

We will now check one by one whether the above mass formula
produces all the required properties of the solution and the
tachyon condensation. At $T=0$, i.e. at the top of the tachyon
potential, $\cos T=1$ and we have from above $M=T_p\left[(N_1 +
{\bar N}_1) + a (N_2 + {\bar N}_2)\right]$, producing the expected
result. It is also easy to see  from (8) that at that point
$\omega^{7-p}$ does not go to zero indicating that the
corresponding solution breaks all the supersymmetry as it should
be. Note that in this case it does not matter whether we have (i)
$N_1 \geq {\bar N}_1$ and $N_2 \geq {\bar N}_2$, (ii) $N_1 \leq
{\bar N}_1$ and $N_2 \leq {\bar N}_2$, (iii) $N_1 \geq {\bar N}_1$
and $N_2 \leq {\bar N}_2$ or (iv) $N_1 \leq {\bar N}_1$ and $N_2
\geq {\bar N}_2$, in all four cases $\omega^{7-p}$ has the same
value. Also at $T=\pi/2$ i.e. at the bottom of the tachyon
potential we get from above $M= T_p\left(|N_1 - {\bar N}_1| + a
|N_2 - {\bar N}_2|\right)$ again producing the expected result.
Here also it does not matter whether we have cases (i) to (iv)
above, we always have $\omega^{7-p}$ going to zero, indicating
that we have a supersymmetric configurations. In fact, for (i) we
have intersecting $(N_1-{\bar N}_1)$ D$p$ branes and $(N_2-{\bar
N}_2)$ D$(p-4)$ branes (when none of the inequalities are
saturated), $(N_1-{\bar N}_1)$ D$p$-branes (when the second
inequality is saturated) and $(N_2-{\bar N}_2)$ delocalized
D$(p-4)$-branes (when the first inequality is saturated).
Similarly for case (ii). For case (iii) we have intersecting
$(N_1-{\bar N}_1)$ D$p$ branes and $({\bar N}_2-N_2)$ ${\bar{\rm
D}}(p-4)$ branes (when none of the inequalities are saturated),
$(N_1-{\bar N}_1)$ D$p$-branes (when the second inequality is
saturated) and $({\bar N}_2-N_2)$ delocalized ${\bar {\rm
D}}(p-4)$-branes (when the first inequality is saturated).
Similarly for case (iv).

Now we will discuss in a bit detail how one can recover the
supersymmetric intersecting brane configuration at the end of
tachyon condensation. Let us first discuss the case $N_1 > {\bar
N}_1$ and $N_2 > {\bar N}_2$. Since for $T\to \pi/2$,
$\omega^{7-p} \to 0$, both $H$ and ${\tilde H}$ goes to unity as
can be seen from eq.(2). Also we notice from (12) that both
$\theta_1$ and $\theta_2$ goes to $\infty$. So, if we take a limit
$(\a_1+\b_1) \sinh\theta_1 \to \epsilon_1^{-1}$, $(\a_2+\b_2)
\sinh\theta_2 \to \epsilon_2^{-1}$ and $\omega^{7-p} \to
\epsilon_1 {\bar \omega}_1^{7-p}$ and $\omega^{7-p} \to \epsilon_2
{\bar \omega}_2^{7-p}$ for some dimensionless parameters
$\epsilon_1,\,\epsilon_2 \to 0$, with ${\bar \omega}_1^{7-p}$ and
${\bar \omega}_2^{7-p}$ = finite, then we get from (3) $b = (7-p)
{\bar \omega}_1^{7-p}$ and $c = (7-p) {\bar \omega}_2^{7-p}$. Both
$F_1$ and $F_2$ in eq.(2) would then reduce to some harmonic
functions \bea F_1 &\to& {\bar H}_1 \,\,\,=\,\,\,
1+\frac{{\bar\omega}_1^{7-p}}{r^{7-p}}\nn F_2 &\to& {\bar H}_2
\,\,\,=\,\,\, 1+\frac{{\bar\omega}_2^{7-p}}{r^{7-p}} \eea with
${\bar \omega}_1^{7-p} = b/(7-p)$ and ${\bar \omega}_2^{7-p} =
c/(7-p)$. The corresponding configuration as can be seen from (1)
with eq.(14) is an intersecting BPS configuration of $(N_1-{\bar
N}_1)$ D$p$ branes with $(N_2-{\bar N}_2)$ D$(p-4)$ branes. One
quarter of susy is restored. The other three cases $N_1<{\bar
N}_1$ and $N_2 < {\bar N}_2$, or, $N_1>{\bar N}_1$ and $N_2 <
{\bar N}_2$, or, $N_1 < {\bar N}_1$ and $N_2 > {\bar N}_2$ will be
very similar and we will not repeat the discussion. However when
one of them is equal i.e. for the case $N_1>{\bar N}_1$ (or $N_1 <
{\bar N}_1$) and $N_2={\bar N}_2$, we find from eq.(12) that
$\theta_2=0$ which implies from eq.(2) that $F_2 \to 1$ (because
$H$ and ${\tilde H}$ go to unity). So, in this case we will have
just one harmonic function ${\bar H}_1$ rather than two. The
corresponding solution can be easily checked from (1) to get
reduced to $(N_1-{\bar N}_1)$ D$p$-branes (or $({\bar N}_1-N_1)$
${\bar {\rm D}}p$-branes). Now one half of susy is restored.
Similarly for the other case i.e. when $N_1={\bar N}_1$ and $N_2 >
{\bar N}_2$ (or $N_2 < {\bar N}_2$). The tachyon condensation can
also be seen for the special case of $N_1={\bar N}_1$ and
$N_2={\bar N}_2$. For this case, at the top of the potential
($T=0$) we get from (13) $M=2T_p(N_1+aN_2)$ as expected and the
corresponding configuration breaks all the supersymmetry. However,
at the end of tachyon condensation i.e. at $T=\pi/2$, $M=0$
corresponding to an empty space-time preserving all supersymmetry.
In fact in this case we see from (8) that $\omega^{7-p}$ vanishes
and so ${\bar H}_1$, ${\bar H}_2$ = 1. For this case, all susy is
restored.

It is also easy to check that the mass formula produces the correct
 result when $N_1 = {\bar N}_1=0$
or $N_2 = {\bar N}_2=0$ or both. For the first case, we have from (13)
$M= a(N_2+{\bar N}_2) T_{p-4}/V_4$ at
the top of the potential and $M=a|N_2-{\bar N}_2|T_{p-4}/V_4$ at the bottom.
For the second case we have $M=(N_1+{\bar N}_1)T_p$ at the top
and $M=|N_1-{\bar N}_1|T_p$ at the bottom as expected.
When both $N_1={\bar N}_1=0$
and $N_2 = {\bar N}_2=0$ we get $M=0$ as expected. It is not difficult
to check that the space-time
configuration at the end of tachyon condensation in these three
cases  indeed reduce to BPS
$(N_2-{\bar N}_2)$ delocalized D$(p-4)$-branes
(or $({\bar N}_2 - N_2)$ delocalized ${\bar {\rm D}}(p-4)$-branes),
$(N_1 - {\bar N}_1)$ $Dp$-branes
(or $({\bar N}_1 - N_1)$ ${\bar {\rm D}}p$-branes) or an empty space-time
respectively. The tachyon vev decouples
in all these three cases at the bottom of the tachyon potential as expected.

To summarize, we have interpreted the supergravity solution given
in (1) with the metric having the isometry ISO($p-4,1$) $\times$
SO(4) $\times$ SO($9-p$) as the intersecting D$p$-${\bar {\rm
D}}p$ and D$(p-4)$-${\bar {\rm D}}(p-4)$-brane system. The five
parameters appeared in the supergravity solution were then
naturally interpreted as related to the five microscopic physical
parameters of the system namely, the number ($N_1$) of
D$p$-branes, the number (${\bar N}_1$) of ${\bar {\rm
D}}p$-branes, the number ($N_2$) of D$(p-4)$-branes, the number
(${\bar N}_2$) of ${\bar {\rm D}}(p-4)$-branes and the tachyon vev
$T$. Based on the physical properties of the solution and the
characteristic behavior of tachyon condensation, we have related
the supergravity parameters with the microscopic physical
parameters of the system. We have obtained the ADM mass
representing the total energy per unit p-brane volume of the
system from the supergravity configuration and related it with the
microscopic physical parameters using the previously mentioned
relations. We have shown that the ADM mass as well as the solution
produce all the required properties of the tachyon condensation
showing that the proposed relations capture the right picture of
the tachyon condensation of the intersecting brane-antibrane
system in the supergravity or closed string description.

\vs{5}

\noindent {\bf Acknowledgements}

\vs{2}

One of us (SR) would like to thank the members of the
Interdisciplinary Center for Theoretical Study at the University
of Science and Technology of China at Hefei, where part of this
work was done, for warm hospitality. We also acknowledge support
by grants from the Chinese Academy of Sciences and the grants from
the NSF of China with 90303002.


\end{document}